\documentclass[preprint, aps]{revtex4-1}

\usepackage{graphicx}% Include figure files
\usepackage{dcolumn}% Align table columns on decimal point
\usepackage{bm}% bold math

\begin{document}

%\preprint{LHEP/DU12-01}

\title{Experimental Conditions for Determination of the Neutrino Mass Hierarchy With Reactor Antineutrinos}% Force line breaks with \\

\author{M. Y. Pac}
\email{pac@dsu.kr}
\address{Dongshin University, Naju, 58245, Korea} 

\begin{abstract}
This article reports the optimized experimental requirements to determine neutrino mass hierarchy using electron antineutrinos ($\bar{\nu}_e$) generated in a nuclear reactor.
The features of the neutrino mass hierarchy can be extracted from the $|\Delta m^{2}_{31}|$ and $|\Delta m^{2}_{32}|$ oscillations by applying the Fourier sine and cosine transform to the $L/E$ spectrum.  To determine the neutrino mass hierarchy above 90\% probability, the requirements on the energy resolution as a function of the baseline are studied at $\sin^{2} 2\theta_{13}=0.1$. If the energy resolution of the neutrino detector is less than $0.04/ \sqrt{E_{\nu}}$ and the determination probability obtained from Bayes' theorem is above 90\%, the detector needs to be located around 48--53 km from the reactor(s) to measure the energy spectrum of $\bar{\nu}_e$. These results will be helpful for setting up an experiment to determine the neutrino mass hierarchy, which is an important problem in neutrino physics.

\begin{description}
\item[PACS numbers]
13.15.+g, 14.60.Pq, 14.60.Lm, 29.85.+c
\item[Keywords] neutrino mass hierarchy, reactor experiment, Fourier transform

\end{description}
\end{abstract}

\maketitle

%\tableofcontents
\newpage
%\narrowtext
%\twocolumn
\section{Introduction}
%\label{intro}
Since the measurement of the large $\sin^{2}2\theta_{13}$ at RENO, Daya Bay, and Double Chooz, the precise measurement of neutrino mass hierarchy, the sign of $\Delta m^{2}_{32}$, has become the focus in neutrino physics\cite{reno,daya,chooz}. It had been believed that the neutrino mass hierarchy can be determined through long-baseline experiments, mainly using accelerator neutrino beams. Recently, the capability of a reactor neutrino experiment at an intermediate baseline to distinguish normal or inverted hierarchy was reported.

For an intermediate-baseline neutrino experiment, many approaches have been proposed; they can be categorized into the $\chi^{2}$ analysis methods, which are discussed in Refs.\cite{ghoshal, qian, ghoshal2, li, takaesu}, and the Fourier-transform methods\cite{qian, ciuffoli, ciuffoli2}. The $\chi^{2}$ analysis methods based on the newly adopted Bayesian approach utilizes all the available information from experiments, and it is straightforward to incorporate the uncertainties in order to evaluate the sensitivity, providing robust and complementary results in the Fourier-transform methods\cite{qian2}. Although the $\chi^{2}$ analysis methods are attractive and interesting, the Fourier-transform methods are more intuitive. The prominent merit of the Fourier-transform methods are that the mass hierarchy can be extracted without precise knowledge of the reactor antineutrino spectrum, the absolute value of the large $|\Delta m^{2}_{31}|$, and the energy scale of a detector. The Fourier-transform methods were introduced to enhance and visualize the structures of mass hierarchy in the frequency spectrum, as first discussed in Ref. \cite{learned}.

In principle, the mass hierarchy can be determined through precise measurements of $|\Delta m^{2}_{31}|$ and $|\Delta m^{2}_{32}|$. As $|\Delta m^{2}_{21}|$ is very small and is only $\sim 3~\%$ of $|\Delta m^{2}_{31}|$, we have to measure $|\Delta m^{2}_{31}|$ and $|\Delta m^{2}_{32}|$ with a precision much better than $3~\%$. However, $|\Delta m^{2}_{31}|$ and $|\Delta m^{2}_{32}|$ have been measured in many experiments with only $\gg 3\%$ precision\cite{pdg}.

The intermediate baseline based on reactor neutrino experiments has been explored using the precise measurement of distortions of the energy spectrum with negligible matter effect. Learned {\it et al.} proposed a new method to distinguish normal and inverse hierarchy after a Fourier transform of the $L/E$ spectrum of reactor neutrinos\cite{learned}. They pointed out that the Fourier power spectrum has a small but not negligible shoulder next to the main peak, and its relative position could be used to extract the mass hierarchy while a non-zero $\theta_{13}$ is considered.

In this paper, we analyze the sensitivity of medium-baseline reactor antineutrino experiments to the neutrino mass hierarchy for a baseline range of 30--60 km and overall energy resolution, $\delta E / \sqrt{E_{\nu}}$, in the range of 0 to $0.08/\sqrt{E_{\nu}}$ with the Fourier-transform method. The optimal baseline length is estimated based on the expected probability of determination.
\\
\section{Detection of reactor antineutrino}
In a nuclear reactor, antineutrinos are mainly produced via the $\beta$-decay of the fission products of the four types of radioactive isotopes,  $^{235}U, ^{238}U, ^{239}Pu$, and $^{241}Pu$, in the fuel. 
The antineutrino flux having energy $E_\nu$ in MeV with thermal power $P_{th}$ in $\rm GW_{th}$ is given as
\begin{equation}
 \frac{dN}{dE_{\nu}}=\frac{P_{th}}{\sum_{k}f_{k}\epsilon_{k}}\phi(E_{\nu})\times 6.24\times10^{21},
\end{equation}   
where $f_k$ and $\epsilon_k$ are the relative fission contribution and the energy released per fission of isotope $k$, respectively. Further, $\phi(E_\nu)$ is the number of antineutrinos produced per fission and is obtained as follows\cite{vogel, huber}:
\begin{eqnarray}
\phi(E_{\nu})&=&f_{^{235}U}e^{0.870-0.160E_{\nu}-0.091E^{2} _{\nu}} \nonumber \\
&+&f_{^{239}Pu} e^{0.896-0.239E_{\nu}-0.0981E^{2} _{\nu}} \nonumber \\
&+&f_{^{238}U} e^{0.976-0.162E_{\nu}-0.0790E^{2} _{\nu}} \nonumber \\
&+&f_{^{241}Pu} e^{0.793-0.080E_{\nu}-0.1085E^{2} _{\nu}}.
\end{eqnarray}
The antineutrino flux is modulated by neutrino oscillation. The antineutrino survival probability $P_{ee}$ is expressed as
\begin{eqnarray}
P_{ee} = 1&-&\cos^{4}(\theta_{13})\sin^{2}(2\theta_{12})\sin^{2}(\Delta_{21}) \nonumber \\
 &-&\cos^{2}(\theta_{12})\sin^{2}(2\theta_{13})\sin^{2}(\Delta_{31}) \nonumber \\
&-& \sin^{2}(\theta_{12})\sin^{2}(2\theta_{13})\sin^{2}(\Delta_{32}).
\end{eqnarray}
The oscillation phase $\Delta_{ij}$ is defined as
\begin{equation}
\Delta_{ij} \equiv \frac{\Delta m^{2} _{ij}L}{4E_{\nu}}, ~~(\Delta m^{2} _{ij} \equiv m^{2} _{i} - m^{2} _{j})
\end{equation}
with a baseline $L$. As $\Delta_{31}$ and $\Delta_{32}$ appear simultaneously in Eq. (3), the effects of mass hierarchy on $P_{ee}$ are hardly recognized. By using the relation between the squared mass differences,
\begin{equation}
\delta m^{2} _{12}+\delta m^{2} _{23} +\delta m^{2} _{31} =0,
\end{equation}
we rearrange Eq. (3) to eliminate the $\Delta_{32}$ term as follows:
\begin{eqnarray}
P_{ee} &=& 1-\cos^{4}(\theta_{13})\sin^{2}(2\theta_{12})\sin^{2}(\Delta_{21}) \nonumber \\
 &-&\sin^{2}(2\theta_{13})\sin^{2}(\Delta_{31}) \nonumber \\
&-& \sin^{2}(\theta_{12})\sin^{2}(2\theta_{13})\sin^{2}(\Delta_{21})\cos(2|\Delta_{31}|)\nonumber \\
&\pm&\frac{\sin^{2}(\theta_{12})}{2}\sin^{2}(2\theta_{13})\sin(2\Delta_{21})\sin(2|\Delta_{31}|).
\end{eqnarray}
The plus (minus) sign in the fifth term on the right-hand side of Eq. (6) corresponds to the normal (inverted) mass hierarchy or NH (IH) in short.

In ongoing reactor experiments, we assume that protons will be used as targets to detect electron antineutrinos via the inverse-beta-decay (IBD), which produces a neutron and positron. The antineutrino distribution observed with a detector having $N_p$ free protons can be expressed for an exposure time $T$ as follows:
\begin{equation}
\frac{dN^{osc}}{dE_{\nu}}=\frac{N_{P}T}{4\pi L^{2}}\frac{dN}{dE_{\nu}}P_{ee}(L, E_{\nu})\sigma_{IBD}(E_{\nu}),
\end{equation}
where $\sigma_{IBD}$ is the cross section of the IBD process and $L$ is the baseline length. 

We use the distribution of the expected antineutrino events from the above expression.
For the IBD cross section, we use the following expression from Vogel and Beacom's work\cite{vogel2}:
\begin{equation}
\sigma_{IBD}=0.0952(E_{e}p_{e})\times 10^{-42} {\rm cm^{2}}.
\end{equation}

In order to study the sensitivity of the mass hierarchy, we use Fourier-transform method together with Monte--Carlo simulations to compare the simulated IBD energy spectrum with the expected spectrum in both the NH and IH cases.

Taking into account the detector response, the reactor electron antineutrino $\bar{\nu}_e$ $L/E$ spectrum becomes
\begin{equation}
\frac{dN^{osc}}{dE^{obs}_{\nu}}=\int{dE_{\nu} \frac{dN^{osc}}{dE_{\nu}}R(E_{\nu},E^{obs}_{\nu}, \delta E_{\nu})},
\end{equation}
where $E_{\nu}$ is the actual $\bar{\nu}_e$ energy, $E^{obs}_{\nu}$ is the observed $\bar{\nu}_e$ energy with the detector response, $\delta E_{\nu}$ is the energy resolution, and $R(E,E')$ describes the detector response function including effects such as the energy resolution and energy scale. In this study, we take the normalized Gaussian function as the response function:
\begin{equation}
R(E^{obs} _\nu, E_\nu, \delta E_\nu)=\frac{1}{\sqrt{2\pi}\delta E_\nu} exp\left\{ {-\frac{(E^{obs}_\nu-E_\nu)^{2}}{2 {\delta E_\nu}^{2}}} \right\}.
\end{equation}
As the neutrino energy is usually measured using scintillators, the energy is typically proportional to the number of photoelectrons, and the error is dominated by the photoelectron statistics.
Therefore, the neutrino energy resolution is proportional to $1/\sqrt{{E}_{\nu}}$. 
In general, the detector energy resolution is parameterized into two parts:
\begin{equation}
\frac{\delta E_\nu}{E_\nu}=\sqrt{ \frac{a^{2}}{E_\nu}+b^{2}}.
\end{equation}
The first term represents the uncertainty from statistical fluctuation, and the second term originates from the systematic uncertainty. In this study, $b=0$ is assumed for simplicity.

\section{Extraction of the mass hierarchy}
Before the measurement of the surprisingly large $\sin^2 2\theta_{13}$, it had been known that at the oscillation maximum of $\Delta_{12}$, which corresponds to a baseline of approximately 58 km, the sensitivity to the mass hierarchy is maximized at $\sin^2 2\theta_{13}\sim0.02$. As $\sin^2 2\theta_{13}$ is no longer small, the sensitivity to mass hierarchy needs to be explored as a function of the baseline, $L$, and the detector energy resolution, $\delta E_{\nu}/E_{\nu}$.

\begin{figure}[htbp]
\centering
\includegraphics[width=150mm]{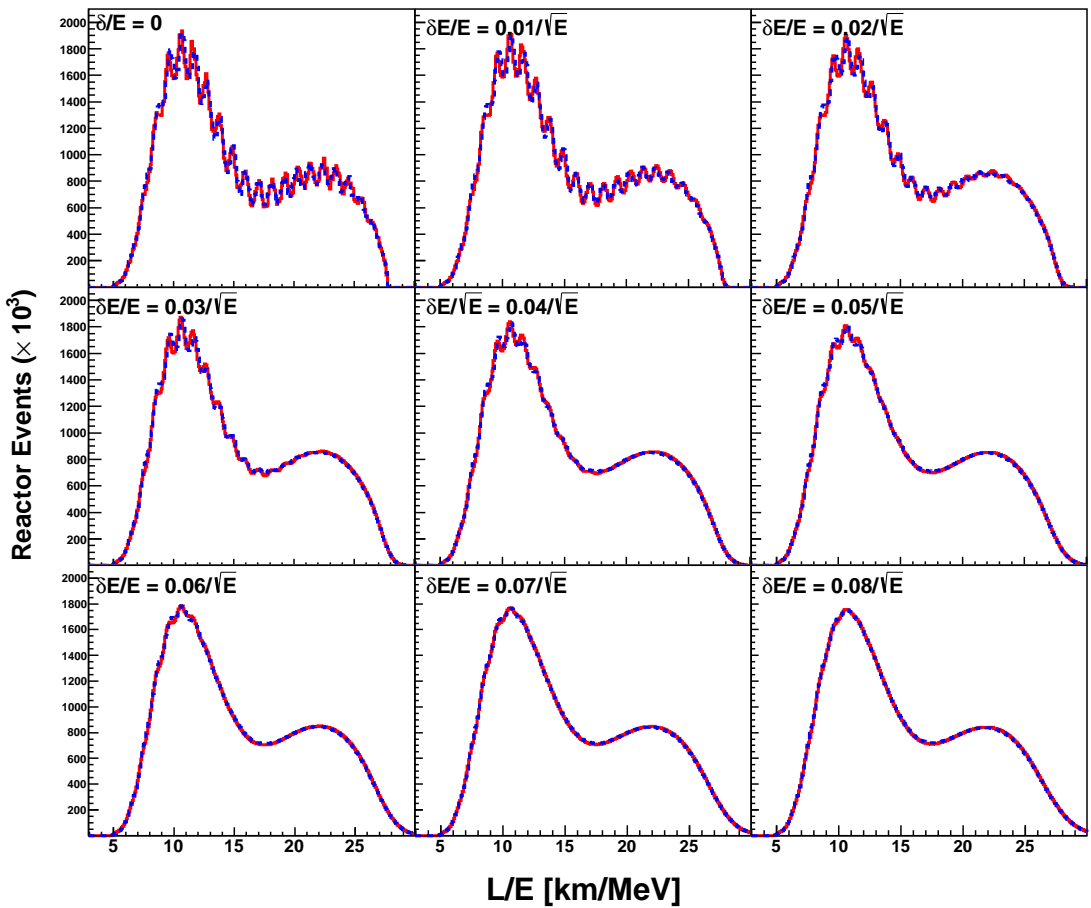}
\caption{\label{fig1}$L/E_{\nu}$ spectra at 50-km baseline for normal hierarchy (solid line) and inverted hierarchy (dotted line) with different detector energy resolutions.}
\end{figure}

In this study, each Monte--Carlo experiment generates a set of 500,000 $\bar{\nu}_e$ events by sampling ${dN^{osc}}/{dE^{obs}_{\nu}}$ with input parameters, $L$ and $\delta E_{\nu}/E_{\nu}$. The default oscillation parameters are taken from Ref. \cite{reno, daya, pdg} and listed in Table \ref{parameter}, together with the explored ranges of baseline and energy resolution.
\begin{table}[htbp]
\caption{\label{parameter}Default values of neutrino oscillation parameters and the explored ranges of other input parameters.}
\begin{tabular}{c c c c}
\hline
$\delta m^{2}_{21}[\rm eV^2]$ & $\Delta m^{2}_{31}[\rm eV^2]$ & $\sin^{2}2\theta_{12}$ &  $\sin^{2}2\theta_{13}$\\ \hline
$7.50\times10^{-5}$ & $2.32\times10^{-3}$ & 0.857 & 0.1\\ \hline \hline

\end{tabular}
%\end{table}

\begin{tabular}{c c c}
$L[km]$ & $a$& b \\ \hline
$30 \le L \le 60$  & $0 \le a \le 0.08$&0\\ \hline
\end{tabular}
%\end{ruledtabular}
\end{table}
A total of 72,000 experiment samples are independently generated for every 2 km in the baseline and every 0.01 of the energy resolution, $\delta E/\sqrt{E_{\nu}}$. Figure \ref{fig1} shows the $\bar{\nu}_e$ $L/E$ spectra at 50-km baseline with the energy resolution varying from 0, which corresponds to an ideal detector, to $0.08/\sqrt{E_{\nu}}$. As all neutrino masses appear in the frequency domain, as indicated by Eq. (6), a Fourier transform of $N(L/E_{\nu})$ would enhance the sensitivity to the mass hierarchy. The frequency spectrum can be obtained using the following Fourier sine transform (FST) and Fourier cosine transform (FCT):
\begin{eqnarray}
FST(\omega)=\int^{t_{max}}_{t_{min}} N(t) \sin(\omega t) dt, \\ \nonumber
FCT(\omega)=\int^{t_{max}}_{t_{min}} N(t) \cos(\omega t) dt,
\end{eqnarray}
where $\omega=2.54 \Delta m^{2}_{ij}$ is the frequency and $t=L/E_{\nu}$ is the variable in $L/E_{\nu}$ space, varying from $t_{min}=L/E_{max}$ to $t_{max}=L/E_{min}$. Once a finite energy resolution is introduced, the phase difference over $L/E_{\nu}$ is significantly smeared out.

Figures 2 and 3 show FST and FCT spectra obtained through the Monte--Carlo simulation from $\delta m^{2}= 0.002$ to 0.028 with energy resolution varied in steps of $2 \times 10^{-5}$. 
The impact of energy resolution is clear because noisy peaks and valleys fluctuate more with increasing magnitude of energy resolution. The main peak and valley are distinctive and can be used to determine the neutrino mass hierarchy while $\delta E_{\nu}/E_{\nu} \leq \sim 0.05/\sqrt{E_{\nu}}$. 

\begin{figure}[htbp]
\centering
\includegraphics[width=150mm]{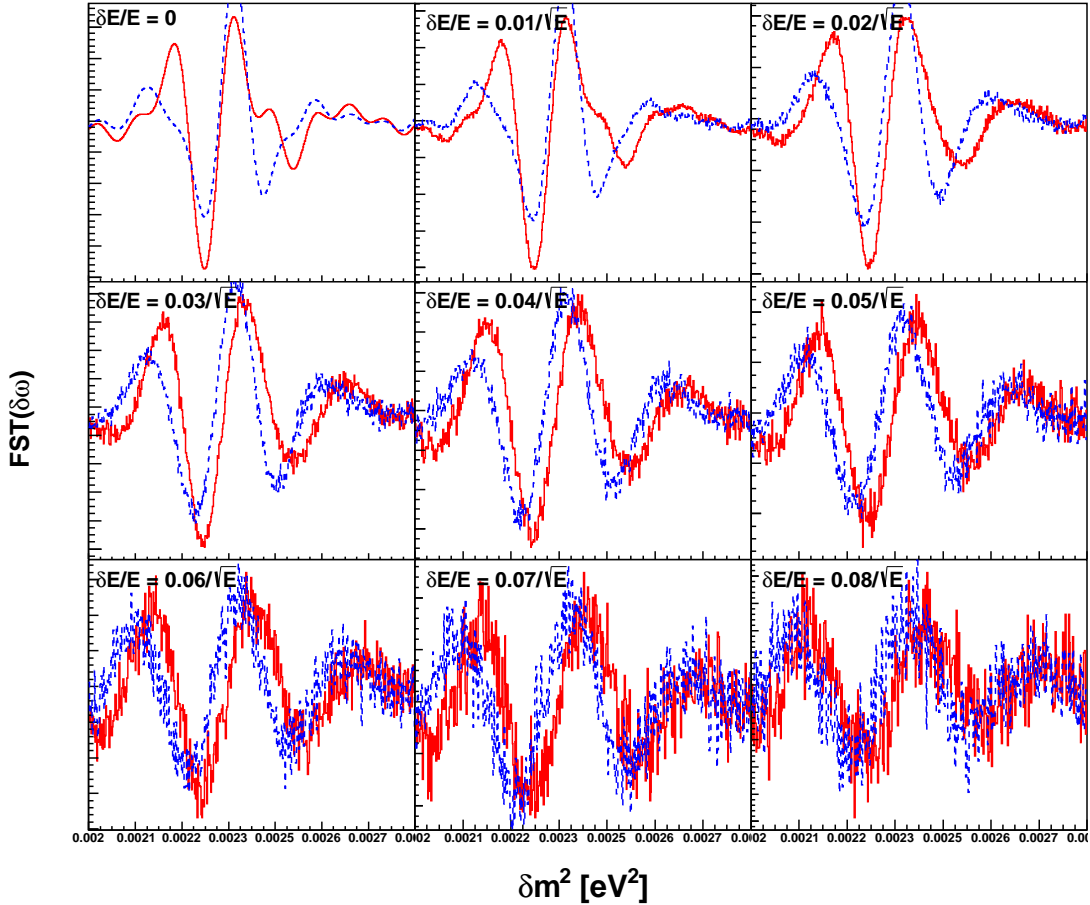}
\caption{\label{fig2}Fourier sine transformed (FST) reactor $\bar{\nu}_{e}$ event rate from 50-km baseline in arbitrary units for $\sin^{2}2\theta_{13}=0.1$, normal hierarchy (solid line), and inverted hierarchy (dotted line) at different energy resolutions.}
\end{figure}
\begin{figure}[htbp]
\centering
\includegraphics[width=150mm]{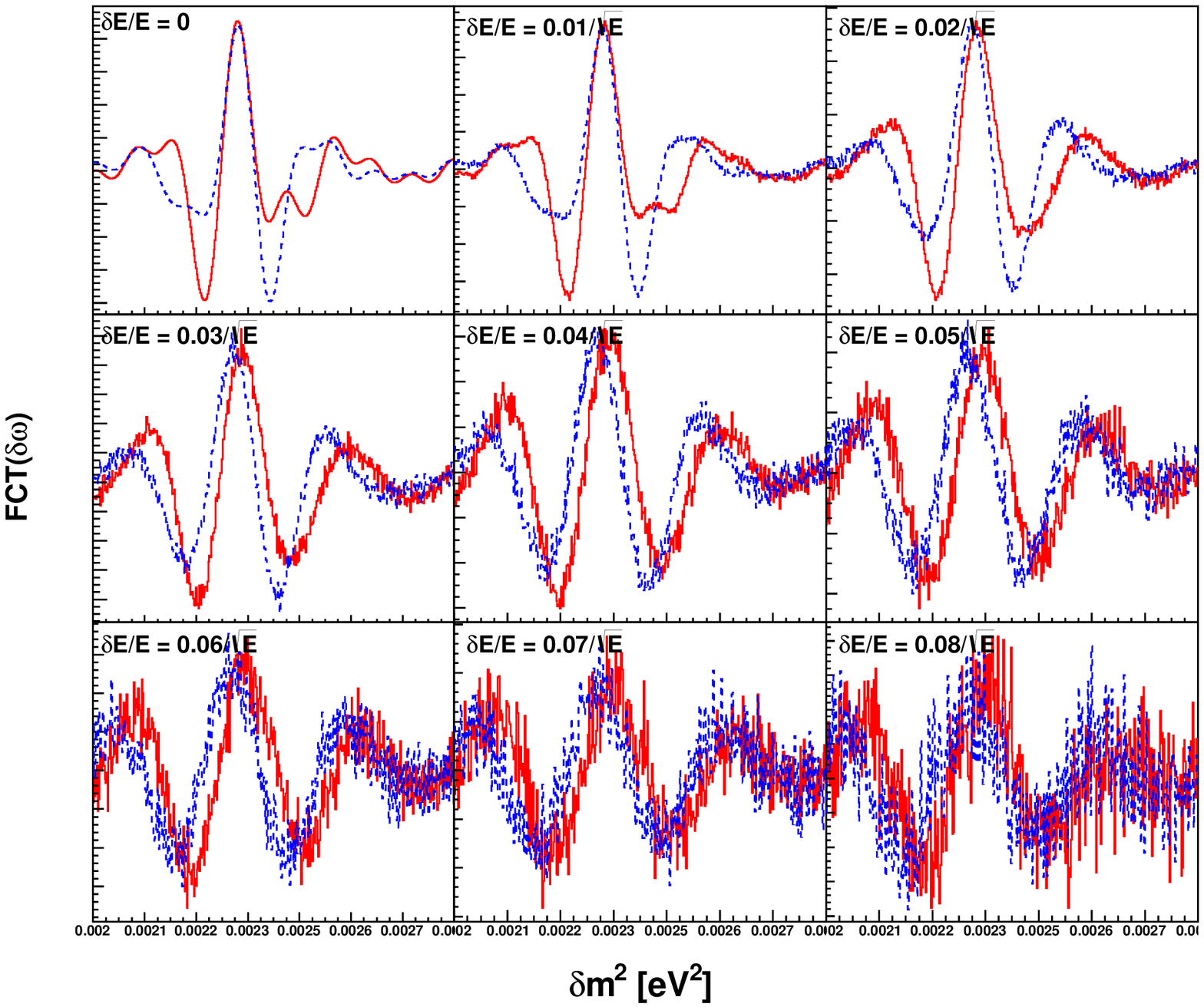}
\caption{\label{fig3}Fourier cosine transformed (FCT) reactor $\bar{\nu}_{e}$ event rate from 50-km baseline in arbitrary units for $\sin^{2}2\theta_{13}=0.1$, normal hierarchy (solid line), and inverted hierarchy (dotted line) at different energy resolutions.}
\end{figure}

We introduce parameters $PV_{FST}$ and $PV_{FCT}$ to quantify the features of FST and FCT spectra:
\begin{equation}
PV_{FST}=\frac{A_p-|A_v|}{A_p+|A_v|}\frac{\delta m^{2}_v - \delta m^{2}_p}{| \delta m^{2}_v - \delta m^{2}_p|}
\end{equation}
and
\begin{equation}
PV_{FCT}=\frac{A_p-|A_v|}{A_p+|A_v|}\frac{\delta m^{2}_v - \delta m^{2}_p}{| \delta m^{2}_v - \delta m^{2}_p|},
\end{equation}
where $A_p$ and $A_v$ are the amplitudes of the peak and valley, respectively, and $\delta m^{2}_p$ and $\delta m^{2}_v$ are the values of $\delta m^{2}$ at the peak and valley positions, respectively. Figure 4 shows the distributions of $PV_{FST}$ and $PV_{FCT}$ for 500 experiments at different energy resolutions. Two clusters of points represented by the red open circle (bottom right) and blue solid circle (top left) in the plane of  ($PV_{FST}$, $PV_{FCT}-PV_{FST}$) corresponding to NH and IH cases, respectively, show their own region exactly when  $\delta E_{\nu}/E_{\nu} \leq \sim 0.05/\sqrt{E_{\nu}}$. The upper and the lower parts of the scatter plot correspond to IH and NH, respectively. It is shown that the distinctive features of NH and IH cases become smeared out as the energy resolution worsens, as shown in Fig. 5.  

At large value of $\sin^{2}2\theta_{13}$, the uncertainty of $|\Delta m^{2}_{31}|$ has a little effect on the FST and FCT spectra. It comes from the fact that $\sin^{2}2\theta_{13}$ is more effective on narrow modulation in the $L/E$ spectrum than $\sin(2|\Delta m^{2}_{31}|)$ does in the last term of Eq. (6). The effect of the uncertainty of $\Delta m^{2}_{31}$ is shown in Fig. 6.

\begin{figure}[htbp]
\centering
\includegraphics[width=150mm]{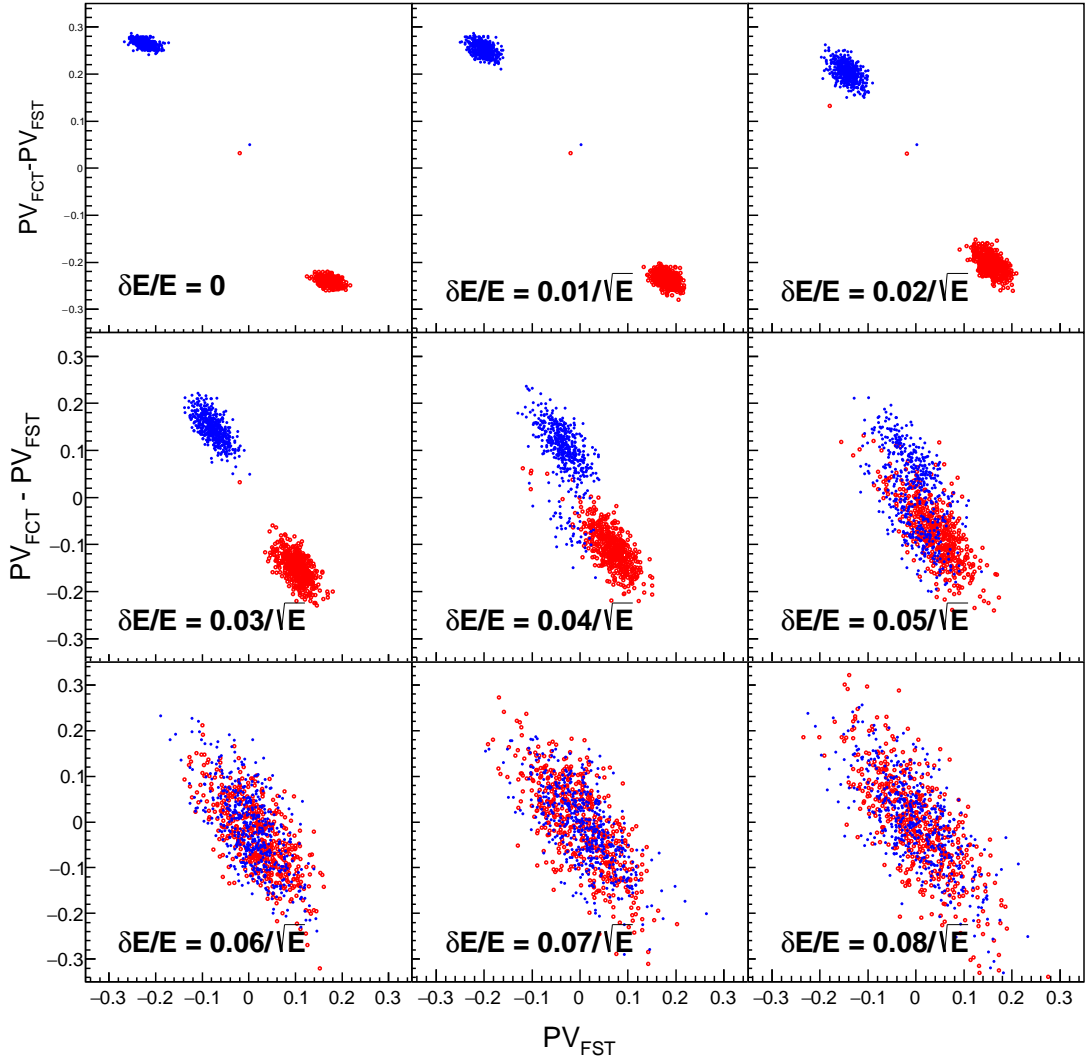}
\caption{\label{fig4}$PV_{FCT}-PV_{FST}$ vs. $PV_{FST}$ scatter plots obtained from 50-km baseline for normal hierarchy (red open circle) and inverted hierarchy (blue solid circle) at different energy resolutions. In the case of an energy resolution of $\delta E / E \le 0.03/\sqrt{E}$, we recognize that points from normal hierarchy (bottom right) and points from inverted hierarchy (top left) are well isolated from each other.}
\end{figure}
\begin{figure}[htbp]
\centering
\includegraphics[width=150mm]{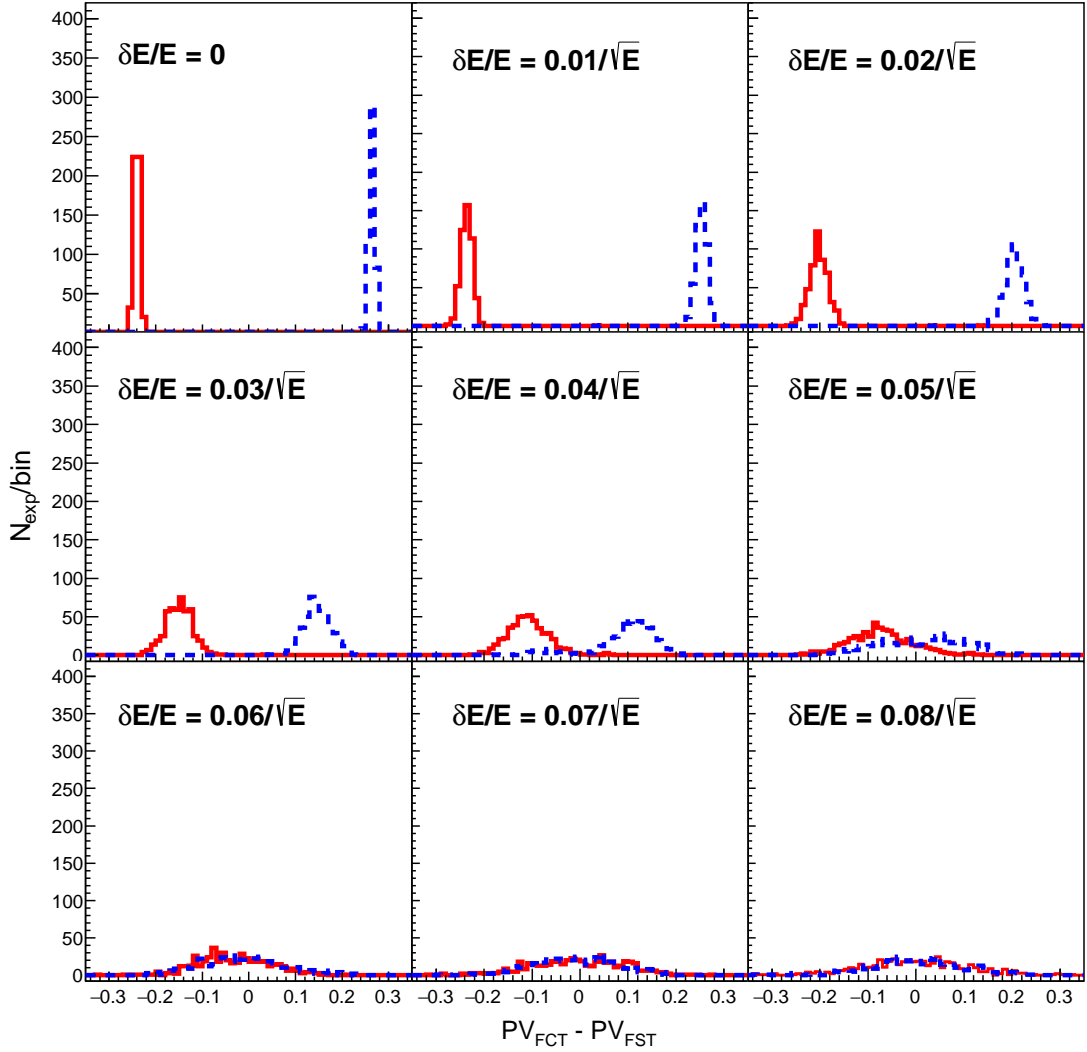}
\caption{\label{fig5}$PV_{FCT}-PV_{FST}$ distributions obtained from 50-km baseline for normal hierarchy (red solid line) and inverted hierarchy (blue dotted line) at different energy resolutions. In the case of an energy resolution of $\delta E / E \ge 0.05/\sqrt{E}$, we could not distinguish normal hierarchy from inverted hierarchy.}
\end{figure}
\begin{figure}[htbp]
\centering
\includegraphics[width=150mm]{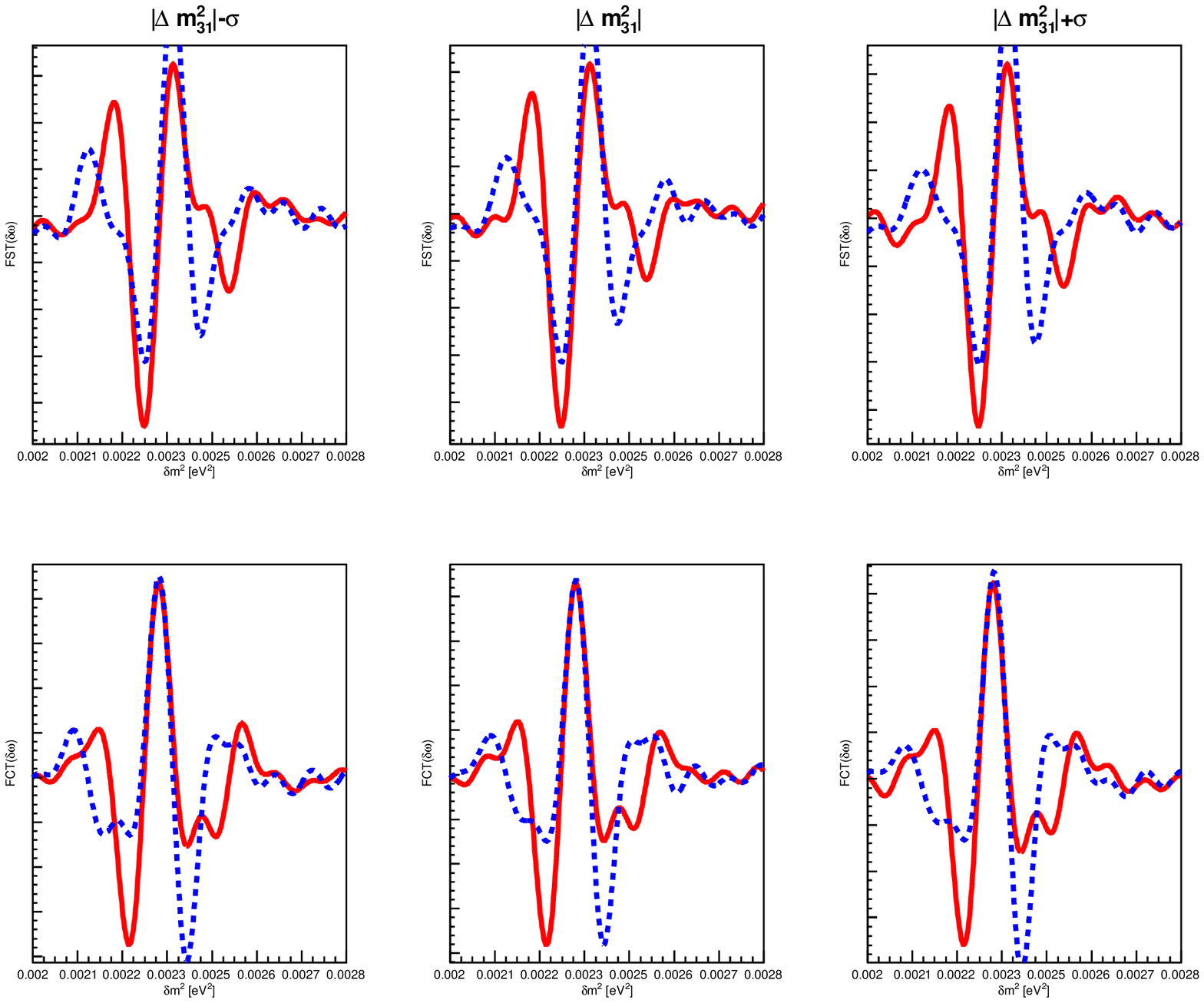}
\caption{\label{fig6}The effect of the uncertainty of $|\Delta m^{2}_{31}|$ on the Fourier spectra at 50-km baseline. Varying $\Delta m^{2}_{31}$ over its uncertainty range has not significant effect on the characteristic features of Fourier sine and cosine spectra. Here  $|\Delta m^{2}_{31}|=2.32^{+0.12}_{-0.09}\times10^{-3}$ is considered\cite{pdg}.}
\end{figure}
Now we consider a method to discriminate between normal and inverted hierarchy using the information we gathered from an experiment.
We will find an experiment on the plane of $PV_{FST}$ versus $PV_{FCT}-PV_{FST}$ as shown in Fig. 4, if we performed the analysis based on an approach suggested in this paper.
The experiment will be placed on the region of NH or IH.
Could we assess quantitatively whether the mass hierarchy of neutrino has normal or inverted hierarchy from the point? 
This is the probability of being NH (IH) given that an experiment  happens to be placed on the NH (IH) region: we name it a success probability, $P_{NH(IH)}$.
The probability is simply calculated using classical Bayes' theorem.  For example, NH concerned, Bayes' theorem says, 
\begin{equation}
P(NH|x)=\frac{P(x|NH)P(NH)}{P(x)},
\end{equation}
where $P(NH|x)$ is the probability of being NH given that an experiment is found on the NH region, $P(x|NH)$ is the probability of being found on the NH reqion given that the hierarchy is NH, $P(NH)$ is the probability of being NH, and $P(x)$ is the probability of being found on the NH region. 

The probability that an experiment will remain in its own region could be calculated from many experiments, which is why we need many experiments. 
According to classical Bayes' theorem, there are
 \begin{equation}
 P_{NH(IH)}=\frac{N^{NH(IH)}_{success}}{N^{NH(IH)}_{total}},
 \end{equation}
 where $N^{NH(IH)}_{success}$ is the number of NH (IH) experiments found in the NH (IH) region and $N^{NH(IH)}_{total}$ is the number of total experiments found in the NH (IH) region. In this approach, 50\% probability implies a null result. 
 
 Figure 7 shows $P_{NH}$ and $P_{IH}$ values obtained from the simulated event samples over the baselines at different energy resolutions. 
We list numerical values of those probabilities acquired from MC samples in Table II.
In the case of $\delta E/ E \le 0.03/\sqrt{E}$, $P_{NH}$ is greater than 95\% for a baseline of 38--56 ${\rm km}$. Similarly, $P_{IH}$ is greater than 95\% at a baseline of 32--52 ${\rm km}$ when $\delta E/E \le 0.04/\sqrt{E}$. As we have no preferred basis to determine which hierarchy is correct, we need to introduce a new probability, which shows that an experiment found in a region remains in its correct region as long as the energy resolution is sufficient. 
 
 The probability that an experiment will be found in its correct region, namely the determination probability, $P_D$, is expressed as 
 \begin{equation}
 P_D \equiv P_{NH} \cdot P_{IH}.
 \end{equation} 
 \begin{figure}[htbp]
\centering
\includegraphics[width=150mm]{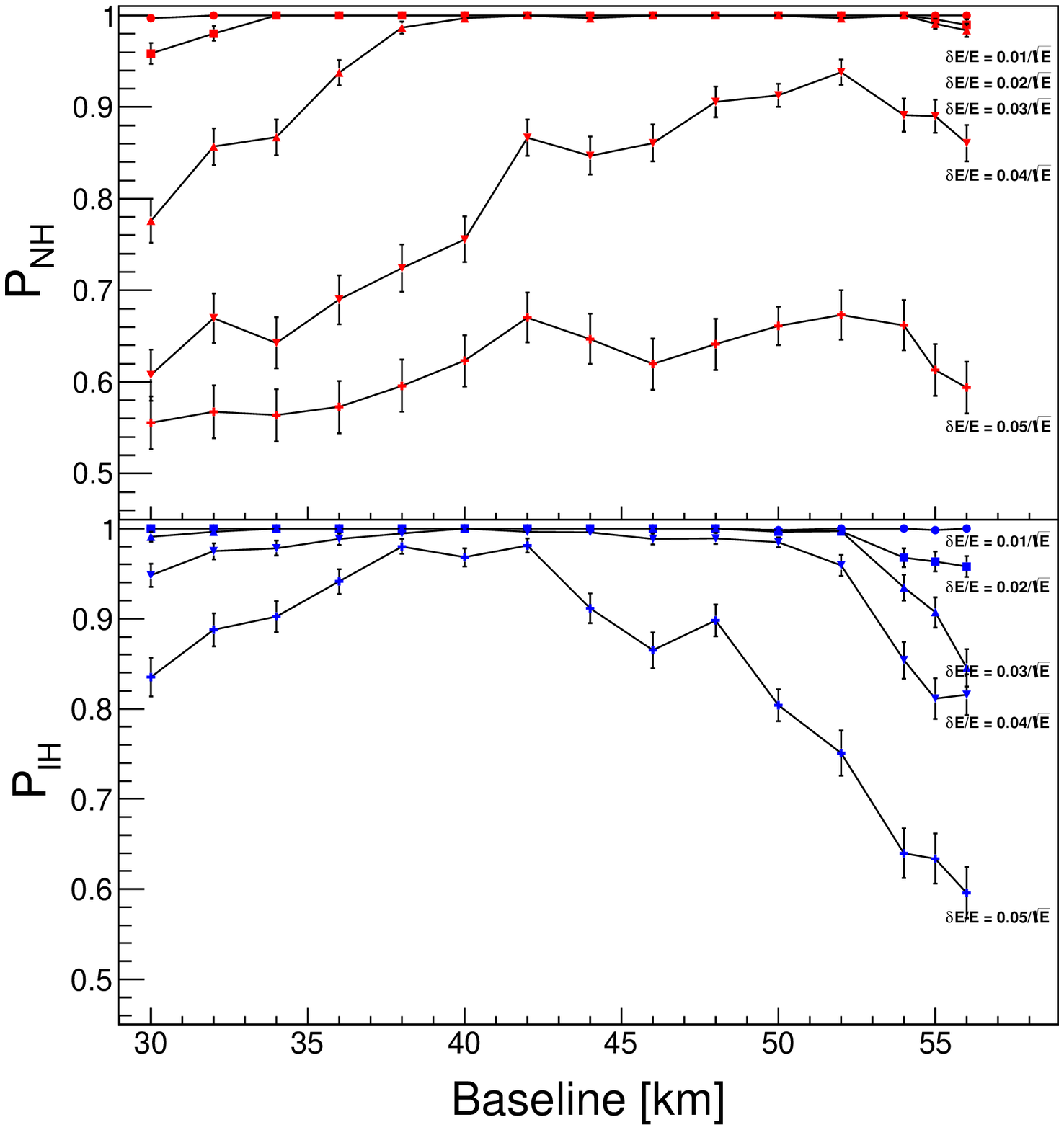}
\caption{\label{fig7}$P_{NH}$ and $P_{IH}$ from Eq. (16) as a function of baseline at different energy resolutions.}
\end{figure}
\begin{figure}[htbp]
\centering
\includegraphics[width=150mm]{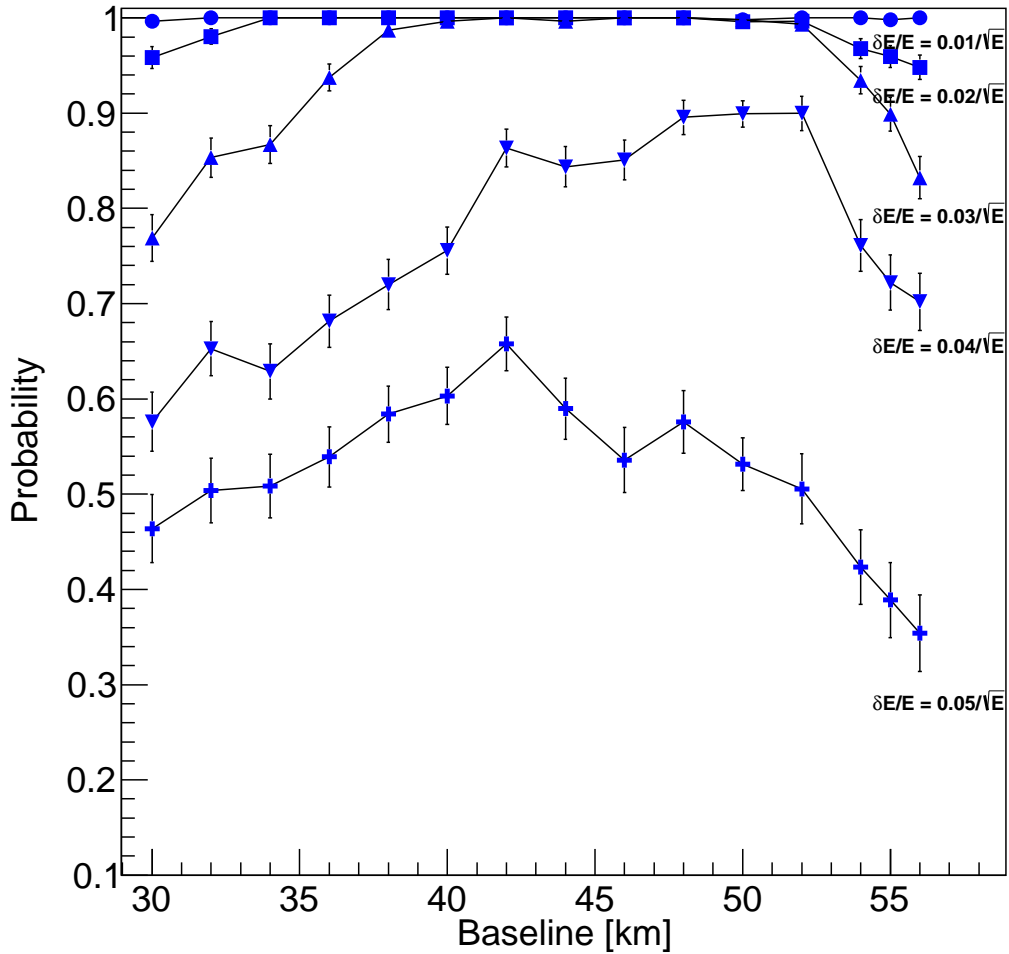}
\caption{\label{fig8}Determination probabilities, $P_D$, from Eq. (17) as a function of baseline at different energy resolutions.}
\end{figure}

As shown in Fig. 8, $P_D$ has a value of $\ge 99\%$ when $\delta E / E \le 0.03/\sqrt{E}$ with a baseline of 40--52 ${\rm km}$. As the energy resolution worsens, $P_D$ rapidly decreases. When $\delta E / E = 0.04/\sqrt{E}$, the baseline is 48--53 ${\rm km}$ at $P_D \ge 90\%$ as shown in Table II.
\begin{table}[htbp]
\caption{\label{prob}Numerical values of $P_{NH}$, $P_{IH}$ and $P_{D}$ for $0.03/\sqrt{E}$ and $0.04/\sqrt{E}$ . Errors are calculated from binomial distribution.}
\begin{tabular}{c | c |c c c c c c}
\hline
baseline [km] & a & $P_{NH}$ & $\sigma_{P_{NH}}$ & $P_{IH}$ & $\sigma_{P_{IH}}$ & $P_{D}$ & $\sigma_{P_{D}}$\\ \hline
30&&0.776& 0.024& 0.991& 0.006& 0.769& 0.025\\
32&&0.857& 0.020& 0.996 &0.004 &0.853& 0.021\\
34&&0.867& 0.020& 1& 0& 0.867& 0.020\\
36&&0.938& 0.014& 1& 0& 0.938& 0.014\\
38&&0.987& 0.007& 1& 0 &0.987& 0.007\\
40&&0.997& 0.003& 1& 0 &0.997& 0.003\\
42&0.03&1& 0& 1& 0& 1& 0\\
44&&0.997& 0.003& 1& 0& 0.997& 0.003\\
46&&1& 0& 1& 0& 1& 0\\
48&&1 &0 &1 &0 &1 &0\\
50&&1 &0& 0.998& 0.002& 0.998& 0.002\\
52&&0.997& 0.003& 0.997& 0.003& 0.993& 0.005\\
54&&1& 0& 0.935& 0.014& 0.935& 0.014\\
55&&0.991& 0.005& 0.907& 0.017& 0.900& 0.018\\
56&&0.984& 0.007& 0.846 &0.021& 0.832& 0.022\\ \hline
30 &  &0.607& 0.0282& 0.948& 0.013& 0.576& 0.031\\
32& &0.670& 0.027& 0.975& 0.009& 0.653& 0.029\\
34& &0.642& 0.028& 0.978& 0.008& 0.629& 0.029\\
36& &0.690& 0.027& 0.988& 0.006& 0.682& 0.027\\
38& &0.724& 0.026& 0.995& 0.004& 0.720& 0.026\\
40& &0.756& 0.025& 1.000& 0.000& 0.756& 0.025\\
42&0.04&0.867& 0.020& 0.996& 0.004& 0.863& 0.020\\
44& &0.847& 0.021& 0.996& 0.004& 0.844& 0.021\\
46& &0.861& 0.020& 0.989& 0.006& 0.851& 0.021\\
48& &0.905& 0.017& 0.989& 0.006& 0.896& 0.018\\
50& &0.913& 0.013& 0.985& 0.005& 0.899& 0.014\\
52& &0.938& 0.014& 0.959& 0.011& 0.900& 0.018\\
54& &0.891& 0.018& 0.854& 0.020& 0.761& 0.028\\
55& &0.900& 0.018& 0.811& 0.023& 0.722& 0.029\\
56& &0.861& 0.020& 0.816& 0.022& 0.702& 0.030\\ \hline
\end{tabular}
\end{table}
\section{Discussion}
We have studied the experimental requirements to determine neutrino mass hierarchy using Fourier sine and cosine transform of the reactor neutrino $L/E$ spectrum at $\sin^{2} 2\theta_{13}=0.1$. The parameters $PV_{FST}$ and $PV_{FCT}$ were defined to extract features of the Fourier sine and cosine spectra, and the mass hierarchy could be obtained from the determination probability, $P_D$ based on Bayes' theorem.

Since the effect of varying $|\Delta m^{2}_{31}|$ over its uncertainty  has little effect on the FST and FCT spectra at $\sin^{2} 2\theta_{13}=0.1$, the $P_D$ is less dependent on the uncertainty of $|\Delta m^{2}_{31}|$ than the value of $\sin^{2}2\theta_{13}$. 

As defined in Eq. (16) and Eq. (17), the probability $P_D$ is closely related to the different features of each mass hierarchy. 
Each value of $P_D$ indicates the probability that an experiment will be found inside its correct $NH$ or $IH$ region on the $PV_{FST}$ and $PV_{FCT}-PV_{FST}$ planes. These different features from different neutrino mass hierarchies suggest that the analysis method described in this paper, which is a simple and straightforward approach, can be used to determine the neutrino mass hierarchy by using the determination probability $P_D$ based on the Fourier sine and cosine transform of the $L/E$ spectrum. 



\section{Acknowledgements}
This work was supported by the National Research Foundation of Korea (NRF-2013R1A1A2011108) and also supported, in part, by the NRF grant No. 2009-0083526.

\end{document}